\documentclass[twocolumn,showpacs,preprintnumbers,amsmath,amssymb,revsymb,prl, superscriptaddress]{revtex4}
\usepackage{graphicx}
\usepackage{mathptmx}

\begin{document}
\title{ Fano Interference between  a STM Tip  and  Mid Gap States in Graphene}


\author{Omid Faizy Namarvar } \email{omid.faizy@nano.cnr.it}
\affiliation{
Universit\'e Grenoble Alpes, Institut NEEL, F-38042 Grenoble, France \\
and CNRS, Institut NEEL, F-38042 Grenoble, France
}
\affiliation{Centro S3, CNR - Istituto Nanoscienze, I-41125 Modena, Italy}
\author{Didier Mayou }  \email{didier.mayou@grenoble.cnrs.fr}
\affiliation{
Universit\'e Grenoble Alpes, Institut NEEL, F-38042 Grenoble, France \\
and CNRS, Institut NEEL, F-38042 Grenoble, France
}

\
\date{\today{}}

\begin{abstract}
We analyze the STM current through  electronic resonances on a substrate as a function of tip-surface distance. We show that when the tip approaches the surface  a Fano hybridization can occur between the electronic resonance on the substrate  and the continuum of conduction  states in the STM tip. A maximum of the density of states of the electronic resonance at some energy  can  then lead to a dip  of the STM signal $dI/dV$. Resonances  in graphene, known as mid gap states,  are good candidates to produce this type of Fano interference. The mid gap states  can be  produced by  local defects or adsorbates and we analyze the cases of  top and hollow configurations of adsorbates.
\end{abstract}


\maketitle

\textit{Introduction} The Scanning Tunneling Microscope (STM) is of major importance in the investigation of  local structural and electronic properties of surfaces or nano-objects.  In particular  resonances  such as  localized states in a nanostructure or near a defect, or  atomic and molecular resonances can well be studied by this technique. Yet a proper interpretation of the STM signal in these cases requires a detailed theoretical analysis due in particular to the complex Fano hybridization effects that can occur  \cite{Fano,Andrey,Plihal,Madhavan,Balatsky,Zawadowski}. 

When  the STM is used in the far distance mode the  signal $dI/dV$ in proportional to the density of states (DOS)  on the substrate close to the STM tip \cite{Tersoff} and the STM signal in the presence of a narrow resonance can present a  complex Fano-like behavior\cite{Fano,Andrey,Plihal,Madhavan,Balatsky,Zawadowski}. This  reflects the combined effect of the coupling of the tip with the different states on the substrate and of  the  hybridization between the localized resonant state on the substrate  and the extended conduction states that are also on the substrate. For example, on a metal,  the coupling of the tip with the conduction states can be the dominant term  in which case the Kondo resonance can lead to a dip of the signal dI/dV  \cite{Zawadowski}. For graphene, due to the low density of  conduction states  near the Dirac energy, the coupling with the localized atomic orbital can be essential. In that case the measured  STM signal dI/dV can present a resonance similar to the density of states on the localized orbital  \cite{Neto,wehling} .

Yet the STM can also be used in a  near contact regime in which  case the coupling between the tip and the substrate is not small. This may modify the electronic structure  of the resonance and increase its width due to the coupling with the continuum of states of the STM tip  \cite{Plihal}. In the case of a magnetic atom  this coupling can even affect the magnetic moment carried by the orbital\cite{Neto}. In addition the standard  Tersoff and Hamann theory  \cite{Tersoff} which assumes that  the current is due to  a weak  tunneling process  between the tip and the substrate is no more valid.  In fact even without a resonance the proximity of the tip with the substrate can deeply modify the STM signal  \cite{Ferrer, Hofer, Blanco,Ryan}. For example it has been shown  \cite{perez} that depending on the tip surface distance the bright spot in the image of graphene can represents either the carbon atoms (far distance) or the centers of the hexagons (short distances). For a resonance on a metallic surface this effect has been theoretically analyzed  \cite{Plihal} in the case of a Kondo impurity  but in the case of graphene this regime has not been considered so far.


In this article  we analyze theoretically the STM current through  a resonant state on a substrate and consider the case where the  tip-surface  varies from a far distance to a near contact regime.  A simple formalism, based on a one channel Landauer model, shows that for small tip-surface distance  the perturbative theory of the STM signal is not valid and the STM image does not represent the local density of states. We find that the high local density of states due to  the resonant state, can lead to  a dip  (anti-resonance)  for  the differential conductance $dI/dV$. This is due to a Fano hybridization  between the electronic resonance on the substrate and the conduction states of the STM tip. Resonances  in graphene, known as mid gap states,  are good candidates to produce this type of Fano interference. The mid gap states  can be  produced by  local defects or adsorbates and we analyze the cases of  top and hollow configurations of adsorbates. 

\begin{figure}[!h]
	\centering 
	\vspace{2mm}
	\scalebox{0.38}{\includegraphics{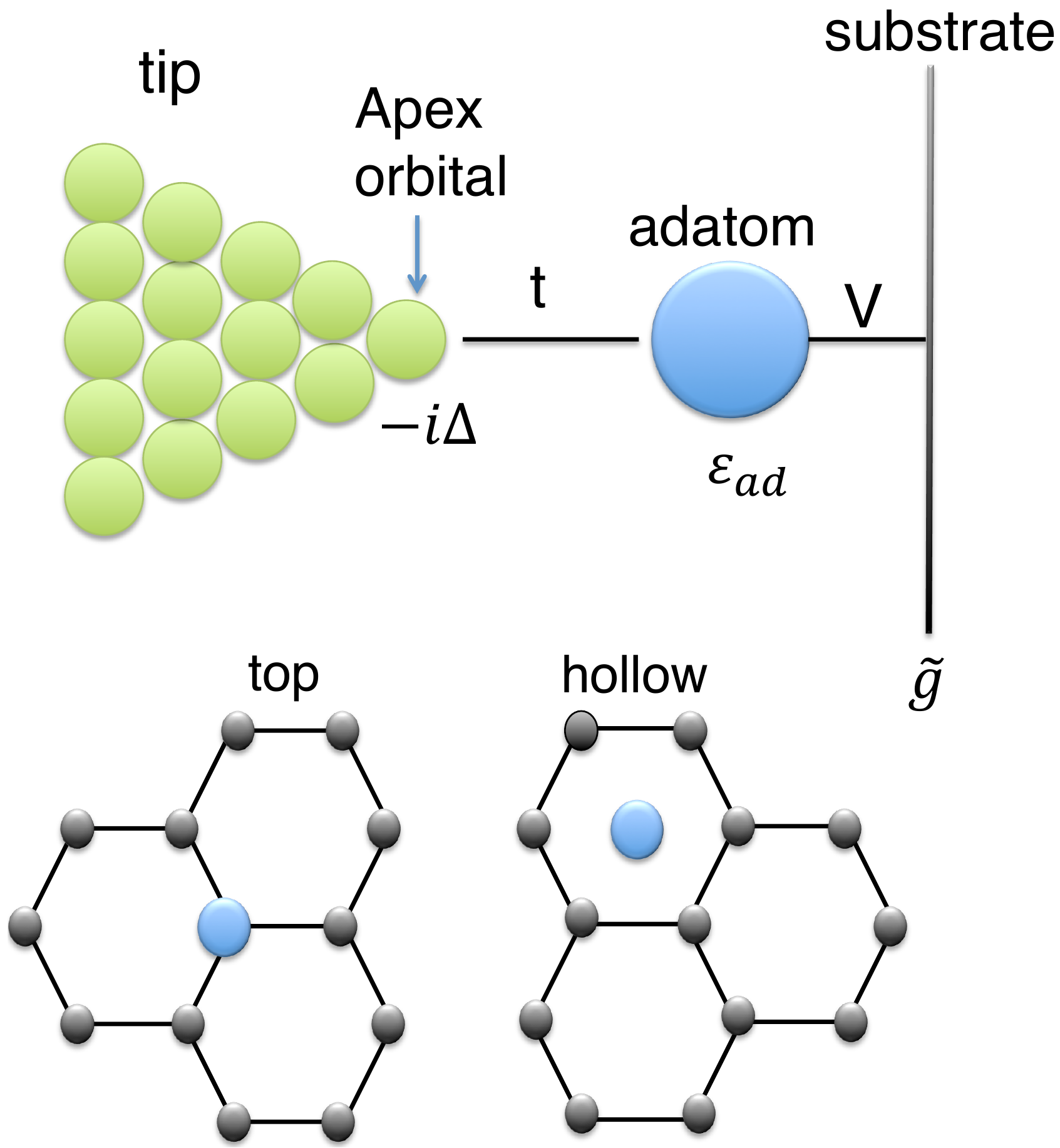}}
	\caption{Upper panel : model of a STM tip approaching an adsorbate. The onsite energy of the apex orbital (AO)  is zero. t is the coupling between AO and the orbital (AD) of the adatom. V is the coupling between AD and the substrate. The states of the substrate that are coupled to the AD are characterized by a Green's function $\tilde{g}(z)$ such that $ \tilde{\Sigma}(z)  = V ^2 \tilde{g}(z)$. Lower panel : left (right) side is the geometry of the top (hollow) position for the adsorbate.}
	\label{fig:model}
	\end{figure}

\textit{Formalism} �  We discuss now the approximations and the formalism used to compute the differential conductance. We consider a simplified model for the tip and assume that the current flows between the tip and the substrate only through one orbital of the tip, that we name the apex orbital (AO).  This implies that the problem of transport can be mapped onto a one channel model as shown for example in \cite{Darancet}.  We define the central part of the circuit as the AO, the left lead is constituted by  the rest of the  STM  tip and  the right lead is constituted by  the zone of the electronic resonance and the rest of the substrate (see figure \ref{fig:model}). Sufficiently far from the apex of the  tip  and from the zone of the electronic resonance in the substrate  the system is assumed to be ballistic.  We thus consider  the formalism developed in \cite{Darancet} which allows to consider the case where the leads are ballistic sufficiently far from the central part of the device but can be non ballistic near the central part of the device. In the one channel case the theory  \cite{Darancet} leads to the formula:

\begin{equation} 
\label{eq:singlefisherdom}
\displaystyle {   T= \tilde{\Gamma}_{\text{STM}} \, \mathcal{G}_{AO}\, \tilde{\Gamma}_{SUB} \ \mathcal{G}_{AO}^{\ast}       }
\end{equation}

where $\tilde{\Gamma}_{\text{STM}}$ and $ \tilde{\Gamma}_{SUB}$ are injection rates for the STM (left) and substrate (right) lead. $\mathcal{G}_{AO}$ is the diagonal element of Green's function on the central part of the device which here is simply the apex orbital. The on-site energy of the AO is taken as the energy origin and set to zero and $  \mathcal{G}_{AO}$  is given by :

\begin{equation} 
\label{eq:greenn}
\displaystyle {   \mathcal{G}_{AO}=\frac{1}{z - \Sigma_{SUB}(z) -\Sigma_{STM}(z) }  } 
\end{equation}

where $\Sigma_{STM}$ and $\Sigma_{SUB}(z) $  are the self-energy of the state $AO$ coupled respectively to the rest of the tip and to the substrate. The expression of $\tilde{\Gamma}_{\text{STM}}$ and $ \tilde{\Gamma}_{SUB}$ (equation (7) of reference \cite{Darancet})  is in general (multi-channel case) different from that of the corresponding terms in the standard Fisher-Lee formula. Yet there is a simplification in the one channel case that was not noted in reference \cite{Darancet}. As we show now  the formula (\ref{eq:singlefisherdom}) is equivalent to the standard Fisher-Lee formula for the one channel case. 


Let us consider the apex orbital (AO) and the right lead (side of the substrate).  According to  \cite{Darancet}  this system can be mapped  on  a one dimensional chain of orbitals $/n>$  where $n$ is an integer. $n=0$ represents the AO orbital and $n\geq 1$ represent all other orbitals of the lead which are states in the substrate. The one dimensional chain has an Hamiltonian given by orbitals on-site energies $a_{n}$ and coupling between successive orbitals $/n>$ and $/n+1>$ which is $b_{n}$. One defines the restricted Green's function $G_{n}(z)$ for site $/n>$ as:

\begin{equation} 
\label{eq:singlefisherdom1}
\displaystyle {    G_{n}(z)=    \frac{ 1}{   z-a_{n} - b_{n}^2G_{n+1}(z) }    }
\end{equation}

$G_{n}(z)$  is the on-site Green's function on site $n$ when all sites $m<n$ are removed. Using the above equation (\ref{eq:singlefisherdom1}) one finds that:

\begin{equation} 
\label{eq:singlefisherdom3}
\displaystyle { \Im (G_{n}(z))= |b_{n}G_{n}(z)|^2 \Im (G_{n+1}(z))}
\end{equation}
 
where $\Im (Z)$ is the imaginary part of complex number $Z$. One can assume that   $a_{n}$ and $b_{n}$ tend to asymptotic values at large $n$. Therefore we set $a_{n}=a$ and $b_{n}=b$ for $n\geq N$ . This means that the effective one dimensional lead is ballistic after some level $n$ with $n\geq N$ as discussed in reference \cite{Darancet} .  $ \tilde{\Gamma}_{SUB}$ is given by equation (7) of reference \cite{Darancet}) in term of the propagator $g_{1,N-1}(z)$ in the lead and of the standard $\Gamma_{SUB}$ . Using  standard formulas with projectors introduced by Zwanzig and Mori  (equation B(12) of \cite{Darancet})  we arrive at the expression  :

\begin{equation} 
\label{eq:singlefisherdom2}
\displaystyle {  g_{1,N-1}(z) =  (b_{1}G_{1}(z)) (b_{2}G_{2}(z)) ... (b_{N-1}G_{N-1}(z))}
\end{equation}

Then  it is straightforward to show that for a one channel model $ \tilde{\Gamma}_{\text{STM}}= \Gamma_{\text{STM}}= -2\Im (\Sigma_{STM}) $ and $ \tilde{\Gamma}_{SUB}= \Gamma_{SUB}= -2\Im (\Sigma_{SUB}(z)) $. The generalized Fisher-Lee formula (equation (\ref{eq:singlefisherdom})) derived in  \cite{Darancet} can therefore be written :

\begin{equation} 
\label{eq:singlefisher} 
\displaystyle {   T(E)= \frac{4 \Im \Sigma_{STM}(E)\Im \Sigma_{SUB}(E)}{|E - \Sigma_{SUB}(E) -\Sigma_{STM}(E)| ^2} }
\end{equation}

Thus  even if the propagation in the tip and in the substrate is not ballistic locally in the vicinity of  the apex orbital  the Fisher-Lee formula (which assumes that the propagation is everywhere ballistic up to the  apex orbital) still applies. Let us emphasize that it is specific to the one channel model which is applicable here because the current is assumed to pass entirely through one orbital i.e.  the apex orbital. As shown by equation (\ref{eq:singlefisher})  a good model of self-energies for the apex orbital   gives enough information for computing current through the STM. We exploit this now to analyze some general aspects of Fano interference in the case of resonances.

\textit{Conditions for the Occurrence of Fano Interferences} � We model the self-energy  due to the coupling of the AO with the tip by $ \Sigma_{\text{STM}}(z)= -i\Delta$. Here $\Delta >0$ is the  width of the resonance of the DOS  of the apex orbital  of the STM tip alone. For simplicity we consider the case where the electronic resonant state is an orbital of an adsorbate atom, but the conclusions  are more general  as it will appear.  We note $t$ the coupling between the AO  and the adsorbate orbital AD.  We note $x$  the dimensionless quantity  $x=t^2/\Delta^2$. In this work we shall take $\Delta \sim 1$eV and $0<t<1$eV. One has therefore $\Sigma_{SUB}(z)=t^2 \tilde{g}_{\text{ad}}(z) $ where  $\tilde{g}_{\text{ad}}(z) $ is the green's function of the adatom orbital coupled to the substrate alone i.e. without coupling to  the STM tip.

\begin{equation}
 \label{eq:greenfunction}
\displaystyle {  \Sigma_{SUB}(z)=x \Delta^2  \tilde{g}_{\text{ad} }(z)  = \frac{ x \Delta^2  }{   z-\epsilon_{ad} - \tilde{\Sigma}(z) }        }
\end{equation}

where $\epsilon_{ad}$ is the on-site energy of the  orbital. $\tilde{\Sigma}(z)$ is the self-energy of the adsorbate orbital due to its coupling with the substrate. Note that, due to their analytical properties,   self-energies can always be written under the form of the equation ( \ref{eq:greenfunction}). Therefore the conclusions drawn below are applicable to other resonant  states and are not specific to adsorbates. For the present model the equation (\ref{eq:singlefisher}) leads to the expression:

\begin{equation}
 \label{eq:tranmission}
\displaystyle { T(E)=  \frac{4\pi x \Delta \tilde{n}_{ad}(E) }{|1+ ix \Delta \tilde{g}_{ad}(E) -i E/\Delta  |^2 } }
\end{equation}

where $\tilde{n}_{ad}(E)=-(1/\pi) \Im(\tilde{g}_{ad}(E+i\epsilon) $ is the density of states on the orbital of the adsorbate without coupling to the STM tip. 

When  the hopping integral $t$ tends to zero, i.e. when the STM tip is at a sufficiently  large distance of the surface,  $ x |\Delta  \tilde{g}_{ad} | \ll1$ ($x=t^2/\Delta^2$). Then, assuming that the energy is within the resonance of the STM tip (i.e. $E/\Delta \ll1$) one get from equation ( \ref{eq:tranmission})  $T(E) \simeq   4\pi x \Delta \tilde{n}_{ad}(E)$. The transmission is proportional to the local density of states close to the apex of the  tip  in agreement with the standard Tersoff-Hamann theory.

Let us examine the effect of the term $ x \Delta  \tilde{g}_{ad}(E) $ in equation (\ref{eq:tranmission}). In order that this term play a role its modulus  must be at least  of the order of $1$. This condition ($   x |\Delta \tilde{g}_{ad}(E) | \geq1$) can be attained.  Indeed for energies close to the resonance we have  $ |\Delta \tilde{g}_{ad}(E)| \sim \Delta /\Delta_R $ where $\Delta_R$ is the  width of the resonance.  $ x|\Delta \tilde{g}_{ad}(E) | \geq 1 $ can be satisfied if $ t \geq t_c$ where the critical valuet $t_c$ is :

\begin{equation}
 \label{tc}
t_c= \sqrt{\Delta \Delta_R} 
 \end{equation}
 
Note that  the transmission  in this regime is necessarily small. Indeed $\pi \tilde{n}_{ad}(E)$ is the imaginary part of $\tilde{g}_{ad}(E) $ therefore $ | \pi  \tilde{n}_{ad}(E)  | < | \tilde{g}_{ad}(E) |  $, so that  after equation (\ref{eq:tranmission}) the transmission is $T(E) \leqslant 4/(x |\Delta  \tilde{g}_{ad}(E)) |  \ll1 $. For $1 \ll x |\Delta  \tilde{g}_{ad}(E))|$ and  for energies $E$ such that  $E \ll \Delta$ equation (\ref{eq:singlefisher}) leads to:

\begin{equation} \label{eq:nono}
\displaystyle {     T(E) \sim -   \frac{4 \Im \tilde{\Sigma}(E)}{x \Delta}      }
\end{equation}

$\tilde{\Sigma}(z)$ can be written as  $ \tilde{\Sigma}(z)  = V ^2 \tilde{g}(z)$   where $V$ is the coupling between the orbital of the adsorbate  and the substrate.  $\tilde{g}(z) $ is the green's function of the states of the substrate which are coupled to the adsorbate orbital. Equation (\ref{eq:nono}) means that the variation with energy of  the transmission $T(E)$ depends only on the effective DOS of the substrate $\tilde{N}(E)=-1/\pi \Im  \tilde{g}(z) $ . In the case of graphene the effective DOS $\tilde{N}(E)$ presents a dip close to zero energy. Therefore the we expect that the STM signal can present a dip instead of a resonance. This is confirmed by the model calculation as shown below. \\

According to formula ( \ref{eq:nono})  the STM tip probes the effective DOS  of the  states of the substrate that are coupled to the tip (via the adsorbate). In fact  the adatom is sufficiently coupled to the STM tip that it can be considered as the apex atom of the STM tip. Finally let us emphasize  that in this regime the transmission $T(E)$  is not  proportional  to the DOS on the adsorbate only coupled to  the substrate  after equation ($ \ref {eq:tranmission}$). $T(E)$  is also not  proportional  to the DOS on the adsorbate coupled to the tip and to the substrate, which value is $\Delta/\pi t^2$.  

We analyze now some models.  The calculation neglects the possibility of collective effects like the Kondo resonance and therefore applies to temperature greater than the Kondo temperature. In addition we do not treat the possible existence of localized magnetic moment on the adsorbate.  This means that the present calculation concerns only one type of spin, and in a true system with localized magnetic moment on the adsorbate the two contributions of the two spins (majority and minority) should be added \cite{Anderson,Neto}. The first two models concern an adsorbate on graphene either in a top position or in a hollow position. For comparison we also consider a model of an adsorbate on a metallic substrate. 
All the effect of the substrate  is contained in the quantity $\tilde{\Sigma}(z)$  which is known  for the three models.

\textit{Adatom in top position on graphene}� In this configuration the adatom is right above one carbon atom of the graphene layer as shown in figure(\ref{fig:model}). The self-energy $\tilde{\Sigma}(z)$ is given by  :

\begin{equation} \label{eq:sigma1}
 \displaystyle {  \widetilde{\Sigma} _{\text{top}} (z) =  \big{(} \frac{V}{D} \big{)}^2  \Big{[} - z  \ln \big{|} 1- \frac{D^2}{z^2} \big{|} \\ -i\pi  \,|z|\, \Theta \big{(} D- |z| \big{)} \Big{]}    }     
\end{equation}

\begin{figure}[htp]

  \centering

  \label{figur}
  \begin{tabular}{cc}
 \includegraphics[width=45mm]{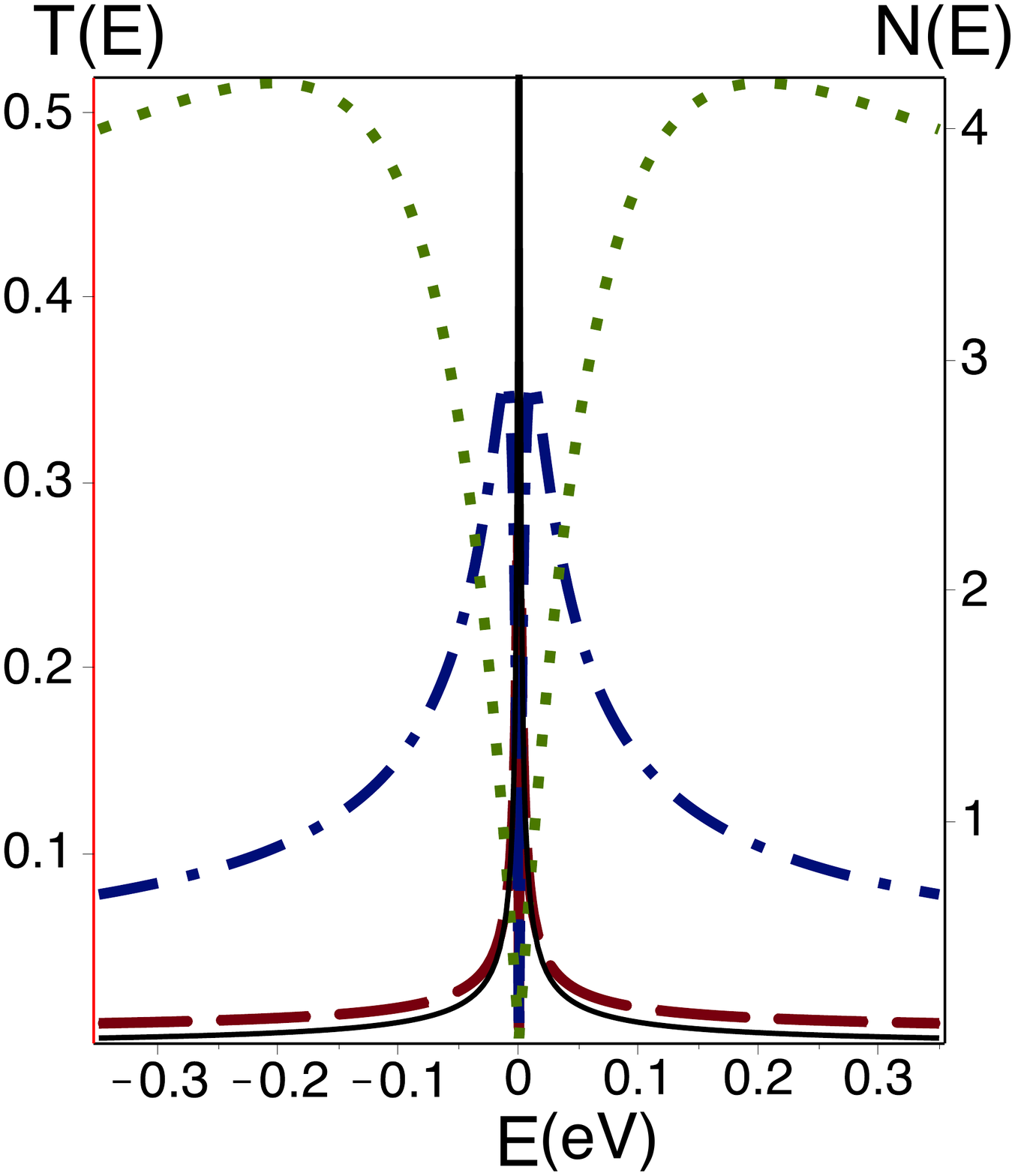}&
 \includegraphics[width=45mm]{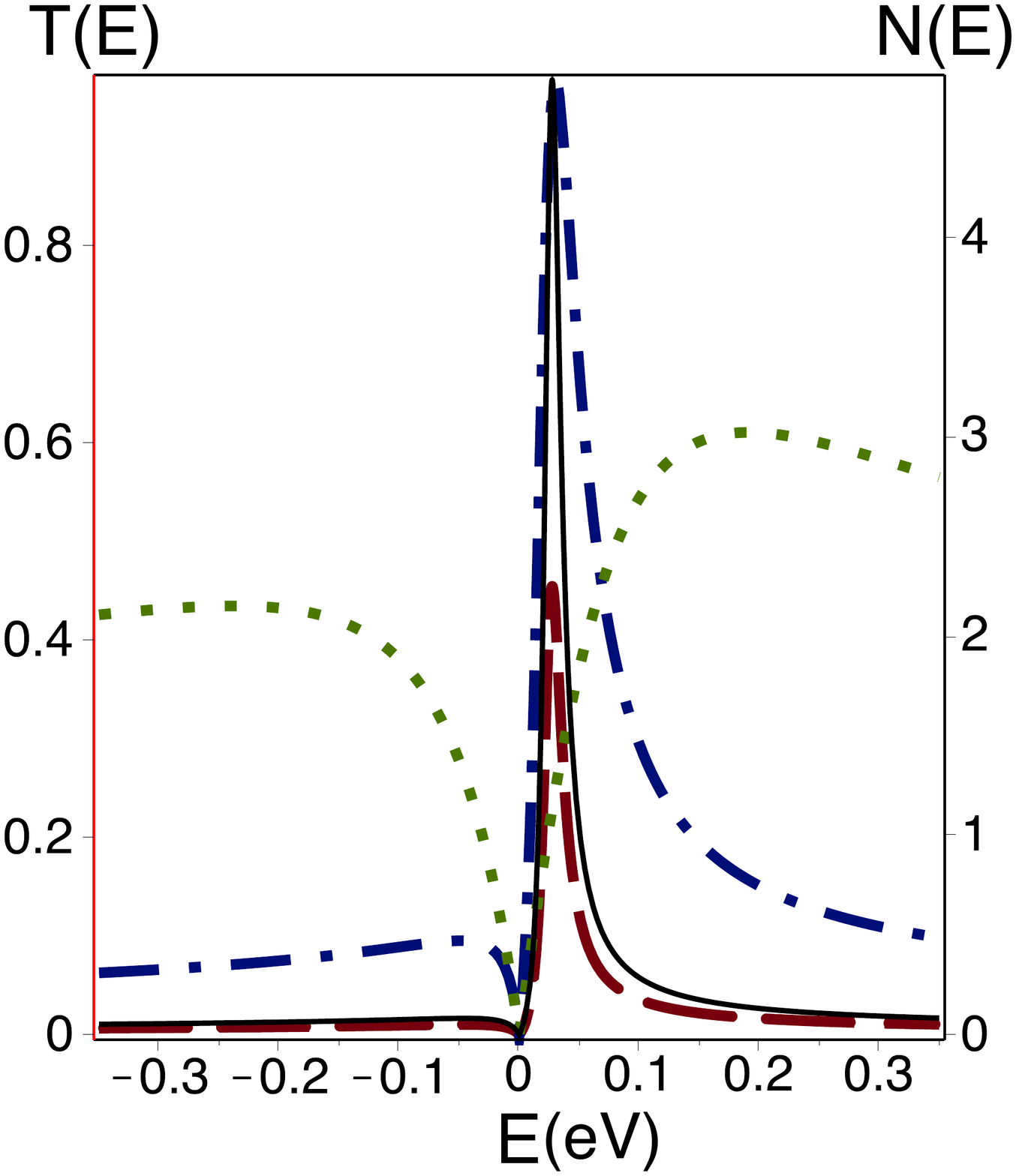}
  \end{tabular}
\caption{Transmission $T(E)$ (left vertical axis) as a function of the energy $E$ for the case of an adatom on top position for different values of coupling $x$ between the apex atom of the tip and the adatom. $x=0.01$ (dashed line)  ,  $x=0.1$ (dashed-dotted line)  and  $ x=1.0$ (dotted line).  Left (right) panel correspond to  on-site energy $\epsilon_{ad}= 0.0$eV ($\epsilon_{ad}= 0.26 eV$). In each panel the density of states N(E) of the adsorbate without STM is represented by the thin solid line. N(E) is given in states/eV on the right vertical axis. }
	
\label{fig:top}
\end{figure}

 $D $  is a high-energy cutoff of order of the graphene bandwidth. $D= \sqrt{\sqrt{3}\pi} t_0 \sim 6$eV where  $t_0\sim 2.8$eV is the hoping energy between nearest neighbors sites of graphene. $V$ is the hybridization amplitude of the orbital of the adatom with the $p_z$ orbital of the nearest  carbon atom \cite{Neto}. Here we consider typical parameters for an hydrogen atom and then $V\sim 5$ eV.  

Figure \ref{fig:top} represents the density of states on the adatom orbital $N(E)$ and the transmission $T(E)$ for two values of the on-site energy of the adatom orbital $\epsilon_{ad}= 0.0$eV  and $\epsilon_{ad}= 0.26$eV . In both cases   the density of states of the adsorbate on the substrate presents a peak at an energy close to the on-site energy $\epsilon_{ad}$. In the symmetric case the peak of the density of states is precisely at the on-site energy $\epsilon_{ad}= 0.0$eV. In the non symmetric case there is a small shift between the position of the peak and  $\epsilon_{ad}= 0.26$eV, due to repulsion of level  by coupling with the continuum of graphene states.  

The transmission is shown in figure \ref{fig:top}  for the two on-site energies  $\epsilon_{ad}$ and different values of $x$. For small $x =0.01$  the transmission varies in accordance with the density of states on the adsorbate $N(E)$ as expected from the Tersoff-Hamann theory. For larger $x=0.1$ the peak of the density $N(E)$ is preserved in the transmission but there is some distortion and $T(E)$ is not strictly proportional to $N(E)$. For $x=1$ the transmission $T(E)$ differs  completely from $N(E)$. In fact close to the energy of the resonance  the coupling  between the resonant state on the graphene side and the conduction state on the tip induce a Fano interference. In this regime $x |\Delta  \tilde{g}_{ad}(E))| \gg 1$ and  the formula (\ref{eq:nono}) is valid. Indeed the transmission $T(E)$ varies like the imaginary part of $\widetilde{\Sigma} _{\text{top}} (E) $ as given by equation (\ref{eq:sigma1}). In this regime $T(E)$ reflects the DOS of the substrate as discussed above.

\textit {Adatom in hollow position on graphene}  �In this configuration the adatom is right above the center of an hexagon of the graphene layer, as shown in figure(\ref{fig:model}). The self-energy $\tilde{\Sigma}(z)$ is given by \cite{Neto}

\begin{align}\label{eq:sigma2}
 \widetilde{\Sigma} _{\text{hollow}} (z) = -z \frac{2}{\pi}  \big{(} \frac{V}{t_0 } \big{)}^2 \Big{ \{} 1 + \frac{z^2}{D^2}   \ln   \big{|}  1- \frac{D^2}{z^2} \big{|}     \Big{\}} \nonumber \\
    -2 i\pi \big{(}\frac{V}{t_0} \big{)}^2 \, \big{|}\frac{z^3}{D ^2} \big{|}\, \Theta \big{(} D- |z| \big{ )}       
\end{align}

where $D$, $t_0$ and  $V\sim 5 $eV have the same values as  for the  case of top position. $V$  is the coupling of the orbital of the adatom with the $p_z$ orbitals of  each of the $6$ nearest neighbors carbon atoms.  It is important to notice that the imaginary part in Equation \ref{eq:sigma2} is much smaller, at low energy $z$,  than for Equation \ref{eq:sigma1} i.e (  $ \displaystyle {   |z|^3/t_0 ^2 \ll  |z|\   }$). This is due to interferences effect in the coupling between the $s$ orbital of the adatom and the $p_z$ orbitals of the six neighbors carbon atoms in the hollow geometry.  This means that an adsorbate in the hollow position is less coupled to low energy states of graphene than in the top position. This will favor  narrower resonance with higher density of states, as shown below.

\begin{figure}[htp]

  \centering

  \label{figur}

  \begin{tabular}{cc}

    \includegraphics[width=40mm]{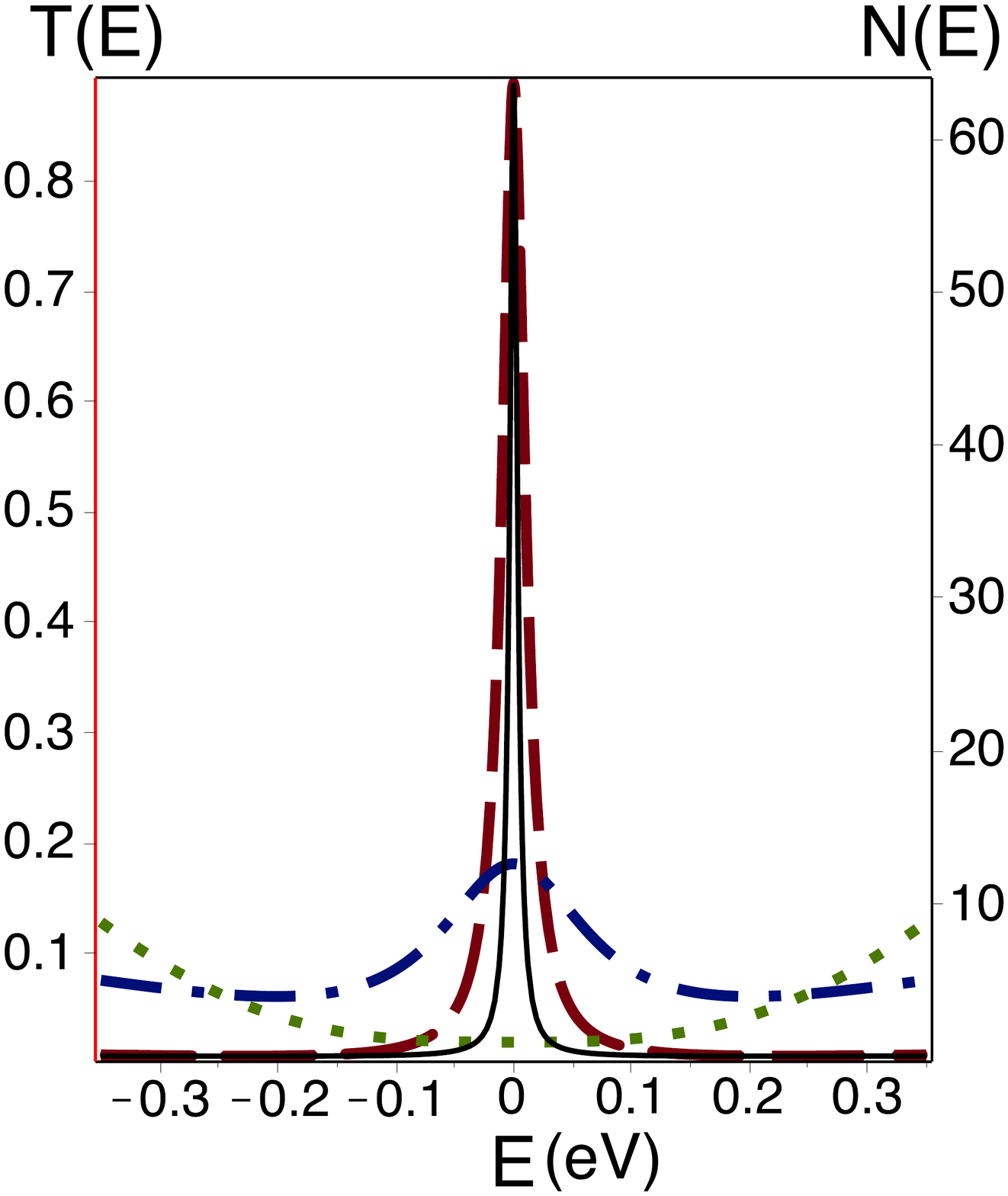}&
    \includegraphics[width=40mm]{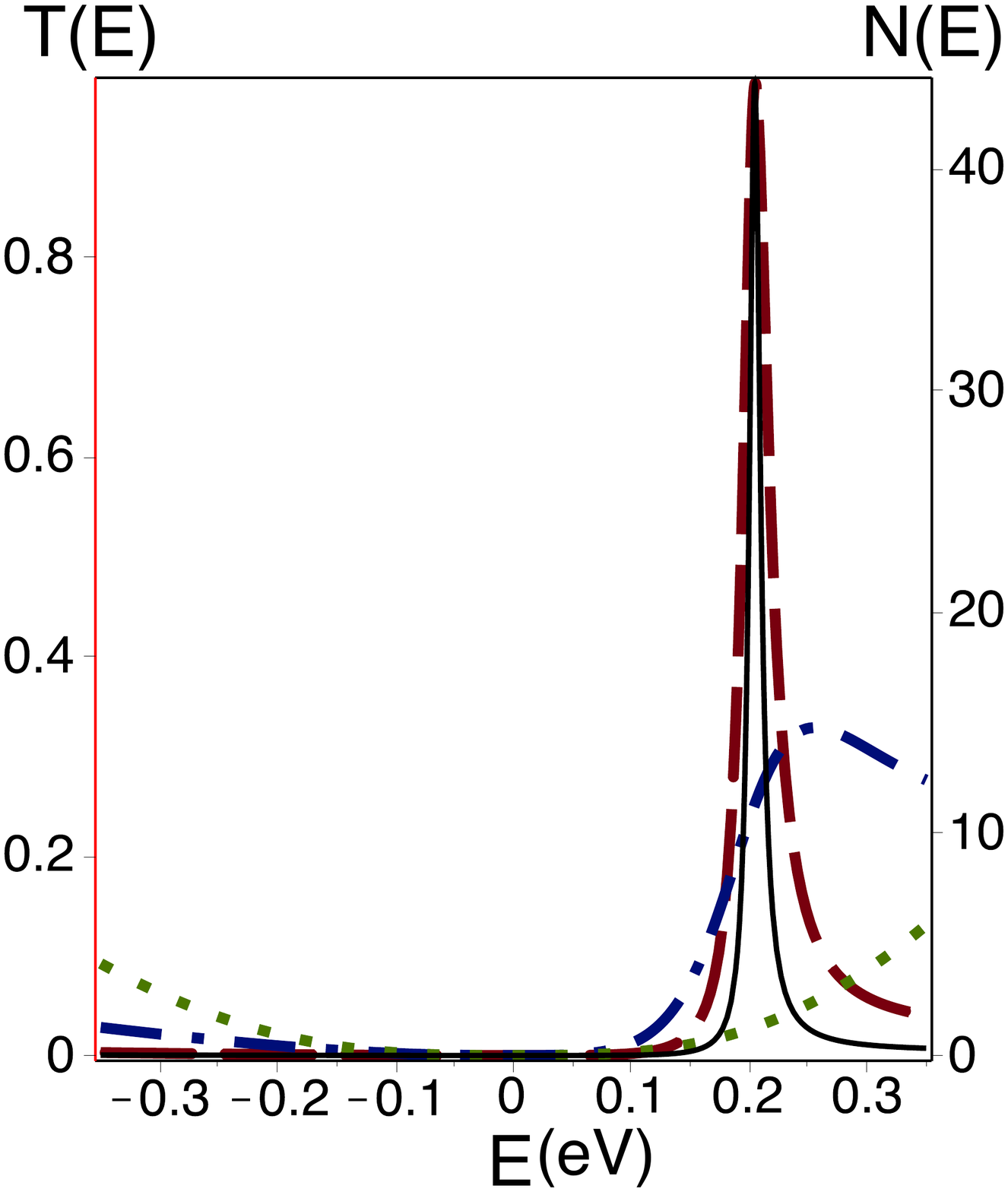}
      \end{tabular}
\caption{ Transmission $T(E)$ (left vertical axis) as a function of the energy $E$ for the case of an adatom on hollow  position for different values of coupling $x$ between the apex atom of the tip and the adatom. $x=0.01$ (dashed line)  ,  $x=0.1$ (dashed-dotted line)  and  $ x=1.0$ (dotted line). Left (right) panel correspond to  on-site energy $\epsilon_{ad}= 0.0$eV ($\epsilon_{ad}= 0.26 eV$). In each panel the density of states N(E) of the adsorbate without STM is represented by the thin solid line. N(E) is given in states/eV on the right vertical axis.   }
\label{fig:hollow}
\end{figure}

Figure \ref{fig:hollow} represents  the density of states of the adsorbate  $N(E)$ and the transmission $T(E)$ for  $\epsilon_{ad}= 0.0$eV  and $\epsilon_{ad}= 0.26$eV . In both cases $N(E)$  presents a peak at an energy close to the on-site energy $\epsilon_{ad}$.  Note that for $\epsilon_{ad} =0$ is a singular case. Indeed $N(E)$ presents a delta peak at $E=0$ with a weight $1/(1+ \frac{2V^2}{\pi t_0 ^2})$. This delta peak is made apparent in Figure \ref{fig:hollow} due to a small  finite imaginary part of $z$, $\Im(z)= 0.005$eV. In the non symmetric case, $\epsilon_{ad}= 0.26$eV,  the shift between the position of the peak and  $\epsilon_{ad}$ is smaller than for the top configuration. This is due to the smaller coupling to graphene states and therefore to a smaller level repulsion effect.

The transmission $T(E)$ is shown in figure \ref{fig:hollow}  for the two values of  $\epsilon_{ad}$ and different $x$. The conclusions are qualitatively similar to those for the top configuration. For small $x =0.01$  the transmission varies in accordance with the density of states on the adsorbate $N(E)$ as expected from the Tersoff-Hamann theory. For larger $x=0.1$ the peak of the density $N(E)$ is preserved in the transmission but there is a strong  distortion and $T(E)$ is not proportional to $N(E)$. For $x=1$ $T(E)$ differs  completely from $N(E)$. For this case the coupling  between the resonant state on the graphene side and the conduction state on the tip induce a Fano interference.  Again  in this regime $x |\Delta  \tilde{g}_{ad}(E))| \gg 1$ and  the formula (\ref{eq:nono}) is valid. The transmission $T(E)$ varies like the imaginary part of $\widetilde{\Sigma} _{\text{top}} (E) $ as given by equation (\ref{eq:sigma1}). In this regime $T(E)$ reflects the DOS of the states of the substrate which are coupled to the adsorbate, as discussed above.

\textit {Adatom on a metallic substrate}---� For comparison we consider a third model that is for  an adsorbate on a metallic substrate. We keep the same value of the coupling orbital $V\sim 5$eV and  we have chosen a rectangular band-model $N(E)=\frac{1}{2W}$ for $-W <E<W$. In the present calculations we take $W=10$eV which is a typical value for a metal. The self-energy $  \widetilde{\Sigma} _{\text{metal}} (z)$ is:

\begin{equation}
\label{eq:sigma3}
\widetilde{\Sigma} _{\text{metal}} (z) = \frac{V^2}{2W} \ln \Big{(}\frac{1+W/z}{1-W/z}  \Big{)}  
\end{equation}

\begin{figure}[htp]
  \centering
  \label{figur}
\includegraphics[width=50mm,height=50mm]{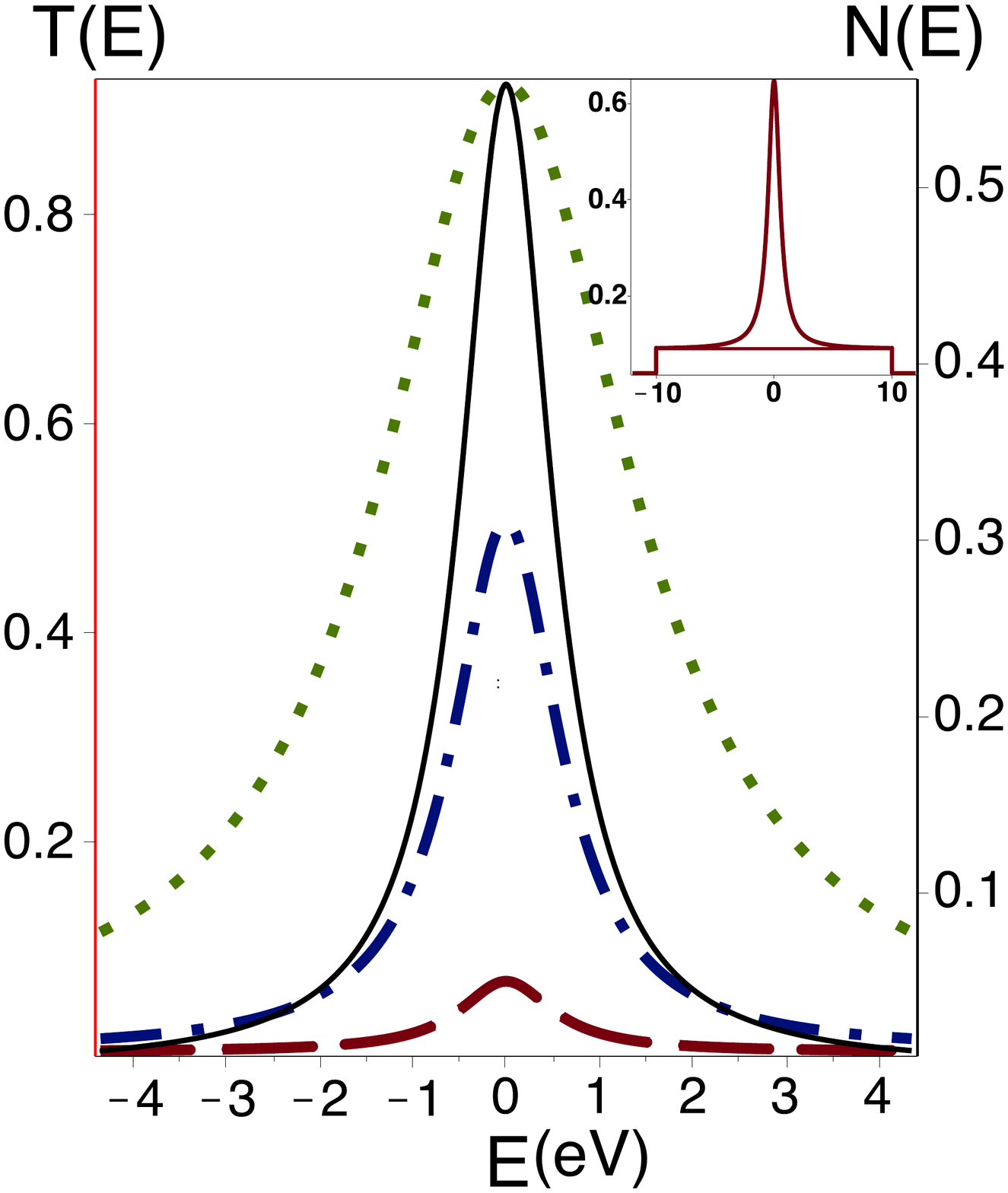}
\caption{  ( Transmission $T(E)$ (left vertical axis) as a function of the energy $E$ for the case of an adatom on a metallic substrate for different values of coupling $x$ between the apex atom of the tip and the adatom. $x=0.01$ (dashed line)  ,  $x=0.1$ (dashed-dotted line)  and  $ x=1.0$ (dotted line). The  on-site energy is $\epsilon_{ad}= 0.0$eV.  The density of states N(E)  of the adsorbate without STM is represented by the thin solid line. N(E) is given in states/eV on the right vertical axis. The insert shows the total density of states of the substrate and the density of state on the adsorbate on the full energy scale of the spectrum. }
\label{fig:metal}
\end{figure}

 In that case the width $\Delta_r$ of the electronic resonance is greater because the density of states of the metallic substrate is higher than that of graphene close to the Dirac energy . Therefore the  condition $t \geq \sqrt{\Delta\Delta_r}$ cannot be achieved for the same values of $\Delta$ and $x$ as before, and the Fano interference plays a minor role. As shown in Figure \ref{fig:metal}  the DOS of the adatom presents a wide resonance on a metallic substrate because the density of state of the metallic substrate is larger than for graphene (top or hollow). As a consequence the phenomena of anomalous STM image does not occur because the resonance is too large to reach to regime $ x |g_{ad}| \gg1 $. Yet this model shows that the characteristics of a resonance on a metallic substrate like its width can be sensitive to the coupling with the tip in a way that cannot be described by the Tersoff-Hamann approach.

\textit{Conclusion}�

To conclude the present work shows that the mid gap states in graphene,  which have received much attention due to their peculiar electronic structure and scattering properties lead not only  to special transport properties \cite{Mayou} but also to special response to STM measurements. We propose a  physical mechanism in which a Fano interference occurs between an electronic resonance on a substrate  and the conduction states of a STM tip. This Fano effect occurs typically if the coupling $t$ between the tip and the resonant state satisfies $t \geq \sqrt{\Delta \Delta_R}$ where $\Delta$ and $\Delta_R$ are the width of the STM resonance  and  the width of the electronic resonance on the substrate. Therefore the  occurence  of this Fano effect is favored for resonances of narrow width  $ \Delta_R$. We expect that this mechanism is rather general but we analyzed specifically two models for resonance on an adsorbate on graphene. These narrow resonances, which are favored by the low density of states in graphene,  have been  studied for realistic parameters of the coupling between the adsorbate and the substrate. We find that the effect should be observable and  is stronger for the hollow position than for the top position. This is because in the hollow position the resonance is more pronounced. Other mid gap states exist in graphene, that are produced for example by vacancies, and they  should lead to similar effects.  Finally we note that Fano effect have been well identified in the context of electronic properties of alloys when the hybridization between localized d orbitals and extended sp orbitals is strong \cite{Mayou2,Nguyen,Pasturel}.



\bibliographystyle{aip}

\end{document}